\documentclass[twocolumn,showpacs]{revtex4}

\usepackage{graphicx}

\begin{document}

\title{State reconstruction of a multimode twin beam using photodetection}

\author{Jan Pe\v{r}ina Jr., \thanks{e-mail:
perinaj@prfnw.upol.cz}}
\affiliation{RCPTM, Joint Laboratory of
Optics of Palack\'{y} University and Institute of Physics of AS
CR, Faculty of Science, Palack\'{y} University, 17. listopadu 12,
771 46 Olomouc, Czech Republic}
\author{Ond\v{r}ej Haderka}
\affiliation{RCPTM, Joint Laboratory of Optics of Palack\'{y}
University and Institute of Physics of AS CR, Faculty of Science,
Palack\'{y} University, 17. listopadu 12, 771 46 Olomouc, Czech
Republic}
\author{V\'{a}clav Mich\'{a}lek}
\affiliation{Institute of Physics of Academy of Sciences of the
Czech Republic, Joint Laboratory of Optics of Palack\'{y} Univ.
and Inst. Phys. AS CR, 17. listopadu 12, 771 46 Olomouc, Czech
Republic}
\author{Martin Hamar}
\affiliation{Institute of Physics of Academy of Sciences of the
Czech Republic, Joint Laboratory of Optics of Palack\'{y} Univ.
and Inst. Phys. AS CR, 17. listopadu 12, 771 46 Olomouc, Czech
Republic}

\begin{abstract}

Robust and reliable method for reconstructing quasi-distributions
of integrated intensities of twin beams generated in spontaneous
parametric down-conversion and entangled in photon numbers is
suggested. It utilizes the first and second photocount moments and
minimizes the declination from experimental photocount histograms.
Qualitatively different forms of quasi-distributions for quantum
and classical states are suggested in the method. The transition
from quantum to classical states caused by an increased detection
noise is discussed. Momentum criterion for non-classicality of
twin beams is suggested.
\end{abstract}

% 42.50.-p    Quantum optics
% 42.65.Lm    Parametric down conversion and production of entangled photons
% 42.50.Ar    Photon statistics and coherence theory


\maketitle

\section{Introduction}

Reconstruction of a quantum state from the measured data
represents one of the most important, and also difficult, problems
in quantum mechanics. This problem has been addressed many times
in the past using different approaches. Mathematicians developed
the method of maximum likelihood \cite{Dempster1977,Vardi1993}
that allows to determine, in principle, a quantum state of an
arbitrary system that suits in the best way to the experimental
data according to a given criterion. However, huge amount of
experimental data needed namely for systems described in larger
Hilbert spaces (including composite systems being in entangled
states like states of twin beams) represents a serious drawback of
this otherwise rigorous method. This limits its useful application
to only very simple quantum systems like those composed of several
two-level systems or spins. Reconstruction of a quantum state of
more complex systems including those characterized by continuous
variables inevitably needs different approaches.

To achieve the preselected level of precision in the
maximum-likelihood method, the number of needed measurement
repetitions is determined by the smallest values contained in a
measured distribution. On the other hand, usually only large
values of the measured distribution contribute significantly to
physical quantities important in the description of the analyzed
system. That is why a smaller number of measurement repetitions
should suffice. Statistical moments of the distribution represent
the tool that allows to separate physically important (and
relevant) information. It is well known that only the first and
second statistical moments of measured physical quantities can be
obtained with sufficient precision after a reasonable number of
measurement repetitions \cite{Perina1991}. On the other hand, the
third and higher statistical moments can scarcely be exploited in
interpreting most experiments in quantum physics.

That is why we have elaborated a method that relies only on the
experimental first and second moments of suitable physical
quantities. However, these moments need not be sufficient for the
characterization of a reconstructed state, especially in case of
entangled quantum states \cite{Bouwmeester2000}. As the example of
twin beams entangled in photon numbers and analyzed below shows
additional parameters are needed for the characterization of such
states. Values of these parameters can then be determined using,
e.g., the most common method that minimizes square declinations
from the experimental histograms.

The presented method relies on the measurement of photocount
statistics. Alternatively, homodyne detection based on mixing the
analyzed field with a coherent local oscillator may be applied
(for review, see \cite{Lvovsky2009}). However, this method is
useful only for fields composed of one spatiotemporal mode or a
small number of modes. It allows to reconstruct even the whole
phase-space quasi-distribution (of field amplitudes) using
qualitatively larger amount of experimental data. That is why only
simpler systems like single-mode fields in Fock states with small
numbers of photons or squeezed two-mode vacuum have been analyzed
by homodyne detection up to now \cite{Lvovsky2009}. On the other
hand, the method presented in this paper allows only for the
reconstruction of quasi-distributions of integrated intensities,
which is sufficient for multi-mode fields. The obtained results
are more reliable as the precision of the proposed method is
superior above homodyne detection.

In this paper, the method for reconstructing a state of twin beams
is presented in Sec.~II. Quasi-distributions of twin beams are
studied in Sec.~III. Conclusions are drawn in Sec.~IV.

\section{Twin beams and their reconstruction}

The analyzed state of twin beams occurring in the process of
spontaneous parametric down-conversion (SPDC) \cite{Mandel1995} is
entangled in photon numbers. In an experiment, histograms of joint
signal-idler (SI) photocount distributions acquired, e.g., by an
iCCD camera are available \cite{Haderka2005a,PerinaJr2012}.
Alternatively, time-multiplexed systems with avalanche photodiodes
\cite{Haderka2004,Rehacek2003}, semiconductor detector arrays
\cite{Ramilli2010}, hybrid photomultipliers
\cite{Bondani2009,Allevi2010} or superconducting bolometers
\cite{Miller2003} can be used to capture photocount histograms.

The states generated in SPDC are highly nonclassical because they
are composed of photon pairs. In more detail, they are
superpositions of states differing in the number of photon pairs.
Each photon in the signal field is accompanied by its twin in the
idler field. As the number of emitted photon pairs is uncertain,
the number of detected signal photons as well as the number of
detected idler photons fluctuate. However, because the state is
entangled in photon numbers, the number of idler photons in an
ideal field equals the number of signal photons \cite{Perina2011}.
In reality, the analyzed optical field contains both photon pairs
and single photons originating either from pairs in which one twin
has not been detected or from straylight. Thus the overall SI
field is found in a general entangled state of a bipartite system
composed of two infinitely large Hilbert spaces.

In quantum theory, this state is described by a joint SI
quasi-distribution $ P(W_s,W_i) $ of integrated intensities $ W_s
$ and $ W_i $ (QDII) of the signal and idler fields, respectively
\cite{Perina1991,Leonhardt1997}. Alternatively, a normal
characteristic function $ C_W $ defined as
\begin{eqnarray}    % 1
 C_W(s_s,s_i) &=& \langle \exp\left[ is_sW_s + is_iW_i\right]
  \rangle_{\cal N} = \nonumber \\
 & & \hspace{-15mm} \int_{0}^{\infty} dW_s \int_{0}^{\infty} dW_i
  P(W_s,W_i) \exp\left[ is_sW_s + is_iW_i\right] \nonumber \\
 & &
 \label{1}
\end{eqnarray}
may be used. This is convenient for statistically independent
fields as their common normal characteristic function $ C_W $
factorizes. An appropriate functional form of the characteristic
function $ C_W $ has to be chosen to describe properly
entanglement in the state. As shown below, the use of the first
and second moments of integrated intensities as parameters is
sufficient provided that we separate the fully entangled (paired)
and noise parts of the state. The overall field is then considered
as composed of three independent components describing photon
pairs, signal noise photons and idler noise photons. Independent
components of signal noise photons and idler noise photons are
assumed to be in the usual multi-mode thermal states
\cite{Perina1991}. The corresponding normal characteristic
functions can be found on the first line of Eq.~(\ref{2}) below.
The form of normal characteristic function appropriate to a
multi-mode paired field (with equally populated pairs of signal
and idler modes) originating in spontaneous process has been
derived in \cite{Perina2005}. It is written on the second line of
Eq.~(\ref{2}) below. The normal characteristic function $ C_W $ of
the overall field can be expressed as
\cite{Perina2005,Perina2006}:
\begin{eqnarray}    % 2
 C_W(s_s,s_i) &=&  \frac{1}{(1-is_sB_s)^{M_s}} \frac{1}{(1-is_iB_i)^{M_i}}
  \nonumber \\
 & & \hspace{-5mm} \mbox{} \times
  \frac{1}{(1-is_sB_p-is_iB_p+s_ss_iB_p)^{M_p}};
\label{2}
\end{eqnarray}
$ M_p $ gives the number of equally-populated modes with the mean
number $ B_p $ of photon pairs per mode. These modes of photon
pairs form the quantum entangled part of the system. On the other
hand, there also exist $ M_s $ ($ M_i $) modes with the mean
number of $ B_s $ ($ B_i $) of single noise signal (idler)
photons. They as noise fields clearly belong to the classical part
of the system. Whereas five independent parameters suffice for the
description of a classical state with a Gaussian form of QDII, six
independent parameters have been introduced for the considered
entangled state. This increase in the number of parameters looks
like a consequence of separation of fully entangled quantum
(paired) and classical parts of the beams. In fact, it reflects a
completely different structure of QDIIs of quantum entangled
states compared to classical ones with a Gaussian shape (see
Sec.~III). We note that the form of normal characteristic function
$ C_W $ in Eq.~(\ref{2}) can be generalized to include also fields
emitted in stimulated parametric down-conversion
\cite{Perina2006}.

In the accompanying experiment, photon pairs have been generated
in non-collinear geometry in type I BBO crystal 5~mm long pumped
by femtosecond pulses coming from the third harmonics of a
Ti:Sapphire laser tuned to 840 nm. Whereas the signal beam has
propagated directly onto the photocathode of an iCCD camera, the
idler beam has been reflected on a dielectric mirror ($ R \approx
99 $~\%) first and then impinged on a different area of the
photocathode. In front of the photocathode, the
nearly-frequency-degenerate signal and idler photons at 280 nm
have been filtered ($ \Delta\lambda_{FWHM} \approx 14$~nm). After
many repetitions of the measurement, the camera has provided a
histogram $ f(m_s,m_i) $ giving the number of measurements
containing the specified numbers of detected signal ($ m_s $) and
idler ($ m_i $) photocounts. The number of measurement repetitions
has allowed to reliably determine from the measured photocount
histogram $ f(m_s,m_i) $ the first and second moments of
photocount statistics denoted as $ \langle m_s \rangle $, $
\langle m_i \rangle $, $ \langle m_s^2 \rangle $, $ \langle m_i^2
\rangle $, and $ \langle m_s m_i \rangle $. Also the first ($
\langle d_s \rangle $, $ \langle d_i \rangle $) and second ($
\langle d^2_s \rangle $, $ \langle d^2_i \rangle $, $ \langle
d_sd_i \rangle $) moments of dark count statistics have been
experimentally obtained. Subsequently, the measured moments of
integrated intensities have been derived according to the
relations eliminating the influence of dark counts:
\begin{eqnarray}   % 3
 \langle {\cal W}_a\rangle_E &=& \langle m_a \rangle - \langle d_a \rangle, \nonumber \\
 \langle (\Delta {\cal W}_a)^2\rangle_E &=& \langle m_a^2 \rangle -
  \langle m_a\rangle^2 - \langle m_a\rangle \nonumber \\
 & & \mbox{} - \langle d_a^2 \rangle + \langle d_a\rangle^2
  + \langle d_a\rangle, \hspace{5mm} a=s,i, \nonumber \\
 \langle \Delta {\cal W}_s \Delta {\cal W}_i\rangle_E &=& \langle m_s m_i \rangle -
   \langle m_s \rangle \langle m_i \rangle \nonumber \\
 & & \mbox{} - \langle d_s d_i \rangle
  + \langle d_s \rangle \langle d_i \rangle;
\label{3}
\end{eqnarray}
$ \Delta {\cal W} = {\cal W} - \langle {\cal W}\rangle_E $.

The signal and idler photons are captured by the photocathode with
non-unit detection efficiencies $ \eta_s $ and $ \eta_i $,
respectively, because of losses in the setup and finite quantum
detection efficiency of the camera. Once the efficiencies $ \eta_s
$ and $ \eta_i $ are known, the moments of integrated intensities
of the fields beyond the crystal can be found. They are related to
the experimental moments of integrated intensities in
Eq.~(\ref{3}) as follows \cite{Perina1991}:
\begin{eqnarray}   % 4
 \eta_a \left[ \langle W_p\rangle + \langle W_a\rangle \right] &=&
  \langle {\cal W}_a\rangle_E , \nonumber \\
 \eta_a^2 \left[ \langle (\Delta W_p)^2 \rangle + \langle (\Delta W_a)^2 \rangle \right] &=&
  \langle (\Delta {\cal W}_a)^2 \rangle_E , \hspace{4mm} a=s,i, \nonumber \\
 \left[ \langle W_p\rangle + \langle (\Delta W_p)^2 \rangle \right] &=&
  \frac{\langle \Delta {\cal W}_s \Delta {\cal W}_i \rangle_E }{\eta_s\eta_i} .
\label{4}
\end{eqnarray}
The last formula in Eq.~(\ref{4}) expresses the fact that only the
field of photon pairs ($ W_p $) creates correlations in the signal
and idler photon numbers. Integrated intensities $ W_s $ and $ W_i
$ of the noise signal and idler fields are not correlated to the
intensity $ W_p $ of the paired field. Equations~(\ref{4})
represent five relations for six independent non-negative moments
$ \langle W_a\rangle $ and $ \langle (\Delta W_a)^2 \rangle $, $
a=p,s,i $. The form of Eq.~(\ref{4}) also shows that bounds for
possible values of these moments exist.

Possible solutions of Eq.~(\ref{4}) form a parametric system that
can be conveniently described by the second moment $ \langle
(\Delta W_p)^2 \rangle $ of the paired field with the allowed
values in the range $ (0,{\rm min}[ \langle (\Delta W_s)^2 \rangle
/ \eta_s^2 , \langle (\Delta W_i)^2 \rangle /\eta_i^2 ] $).
According to Eq.~(\ref{4}), the remaining moments are:
\begin{eqnarray}   % 5
 \langle W_p\rangle &=& \frac{\langle \Delta {\cal W}_s \Delta {\cal W}_i \rangle_E
  }{\eta_s\eta_i} - \langle (\Delta W_p)^2 \rangle , \nonumber \\
 \langle W_a\rangle &=& \frac{\langle {\cal W}_a\rangle_E}{\eta_a} -
  \frac{\langle \Delta {\cal W}_s \Delta {\cal W}_i \rangle_E }{\eta_s\eta_i}
  +\langle (\Delta W_p)^2 \rangle , \nonumber \\
 \langle (\Delta W_a)^2 \rangle &=& \frac{ \langle (\Delta {\cal W}_a)^2
  \rangle_E}{ \eta_a^2} -  \langle (\Delta W_p)^2 \rangle ,
  \hspace{3mm} a=s,i .
\label{5}
\end{eqnarray}

It can be shown that the solution for the relations (\ref{4})
exists only if the following inequality is obeyed:
\begin{eqnarray}    %   6
 \eta_s \hspace{-2mm} &\ge& \hspace{-2mm}
  \frac{ \langle \Delta {\cal W}_s \Delta {\cal W}_i \rangle_E/\alpha -
  {\rm min} \left[ \langle (\Delta {\cal W}_s)^2 \rangle_E,
  \langle (\Delta {\cal W}_i)^2 \rangle_E /\alpha^2 \right] }{
  {\rm min} \left[ \langle {\cal W}_s \rangle_E,
  \langle {\cal W}_i \rangle_E /\alpha \right] },
  \nonumber \\
 & &
\label{6}
\end{eqnarray}
where $ \alpha = \eta_i / \eta_s $ denotes the ratio of quantum
efficiencies. For typical experimental data, the inequality
(\ref{6}) puts a strong requirement to the allowed values of
quantum efficiencies $ \eta_s $ and $ \eta_i $.

Quantum theory of photo-detection \cite{Perina1991} shows that a
joint signal-idler photon-number distribution $ p $ corresponding
to the normal characteristic function $ C_W $ in Eq.~(\ref{2})
takes the form of a two-fold convolution of three Mandel-Rice
distributions \cite{Perina1991}:
\begin{eqnarray}  % 7
 p(n_s,n_i) &=& \sum_{n=0}^{{\rm min}[n_s,n_i]} p(n_s-n;M_s,B_s)
  \nonumber \\
 & & \mbox{} \times
  p(n_i-n;M_i,B_i) p(n;M_p,B_p),
\label{7}
\end{eqnarray}
where $ p(n;M,B) = \Gamma(n+M) / [n!\, \Gamma(M)] B^n/(1+B)^{n+M}
$ and $ \Gamma $ denotes the gamma-function. Mean photon numbers $
B_a $ and numbers $ M_a $ of modes as they were introduced in
Eq.~(\ref{2}) are obtained from the moments written in
Eq.~(\ref{5}):
\begin{eqnarray}   % 8
 B_a = \frac{\langle (\Delta W_a)^2 \rangle }{ \langle W_a
  \rangle } , \hspace{4mm}
 M_a = \frac{ \langle W_a \rangle^2 }{ \langle (\Delta W_a)^2 \rangle
  }, \hspace{4mm} a=p,s,i.
\label{8}
\end{eqnarray}

A joint signal-idler photocount distribution $ p_c $ describing
theoretically the measured photocount histogram $ f $ is then
derived from the photon-number distribution $ p $ in Eq.~(\ref{7})
provided that the detection process is characterized. An iCCD
camera with $ N $ pixels, detection efficiency $ \eta $ and
dark-count rate $ D \equiv \langle d\rangle / N $ is described by
the probabilities $ T(m,n) $ of having $ m $ photocounts out of a
field with $ n $ photons in the form \cite{PerinaJr2012} ($ a=s,i
$):
\begin{eqnarray}     % 9
 T_a(m,n) &=& \left( \begin{array}{c} M \cr m \end{array} \right)
  (1-D_a)^{N_a} (1-\eta_a)^{n} (-1)^{m} \nonumber \\
 & & \hspace{-10mm} \times
  \sum_{l=0}^{m}
  \left( \begin{array}{c} m \cr l \end{array} \right) \frac{(-1)^l}{(1-D_a)^l}
  \left( 1 + \frac{l}{N_a} \frac{\eta_a}{1-\eta_a}
   \right)^{n} .
\label{9}
\end{eqnarray}
The formula in Eq.~(\ref{9}) allows us to express the joint
signal-idler photocount distribution $ p_c $ as follows:
\begin{eqnarray}   % 10
  p_c(m_s,m_i) = \sum_{n_s,n_i=0}^{\infty} T_s(m_s,n_s)
   T_i(m_i,n_i) p(n_s,n_i) .
\label{10}
\end{eqnarray}

The method of least square declinations then minimizes a function
$ {\cal D} $ defined as:
\begin{equation}  % 11
 {\cal D} = \sqrt{ \sum_{m_s,m_i=0}^{\infty} \left[p_c(m_s,m_i) -
  f(m_s,m_i)\right]^2 }.
\label{11}
\end{equation}
In this minimization, all elements of the experimental histogram $
f $ are taken into account.

In the experiment, numbers of detected photocounts are monitored
in three areas on the photocathode of the iCCD camera
\cite{PerinaJr2012}. The first and second areas are illuminated by
the signal and idler fields, respectively, and typically contain
several detection events. The last area monitors noise (straylight
and dark counts). The numbers of photocounts are collected from
typically $ 10^5 $ frames; each frame arises from detection of the
signal and idler fields generated from one pump pulse.
Subsequently, the histogram $ f $ giving the number of frames with
defined numbers of signal and idler photocounts is built and the
first and second moments of photocount numbers are determined. The
knowledge of moments of the noise monitored in the third strip is
used to eliminate the effect of this noise to the moments of the
measured signal and idler photocount numbers. Relations in
Eq.~(\ref{3}) are then applied to derive the moments of
experimental integrated intensities. In a typical experiment
analyzed below, the following values of moments have been found: $
\langle {\cal W}_s \rangle_E = 2.411 $, $ \langle {\cal W}_i
\rangle_E = 2.353 $, $ \langle (\Delta {\cal W}_s)^2 \rangle_E =
0.079 $, $ \langle (\Delta {\cal W}_i)^2 \rangle_E = 0.095 $, and
$ \langle \Delta {\cal W}_s \Delta {\cal W}_i \rangle_E = 0.598 $.
Detection efficiencies $ \eta_s = 24.3 $\% and $ \eta_i = 23.5 $\%
have been obtained in an independent measurement.

\begin{figure}    % fig. 1
 (a)
 \resizebox{0.8\hsize}{!}{\includegraphics[scale=0.26]{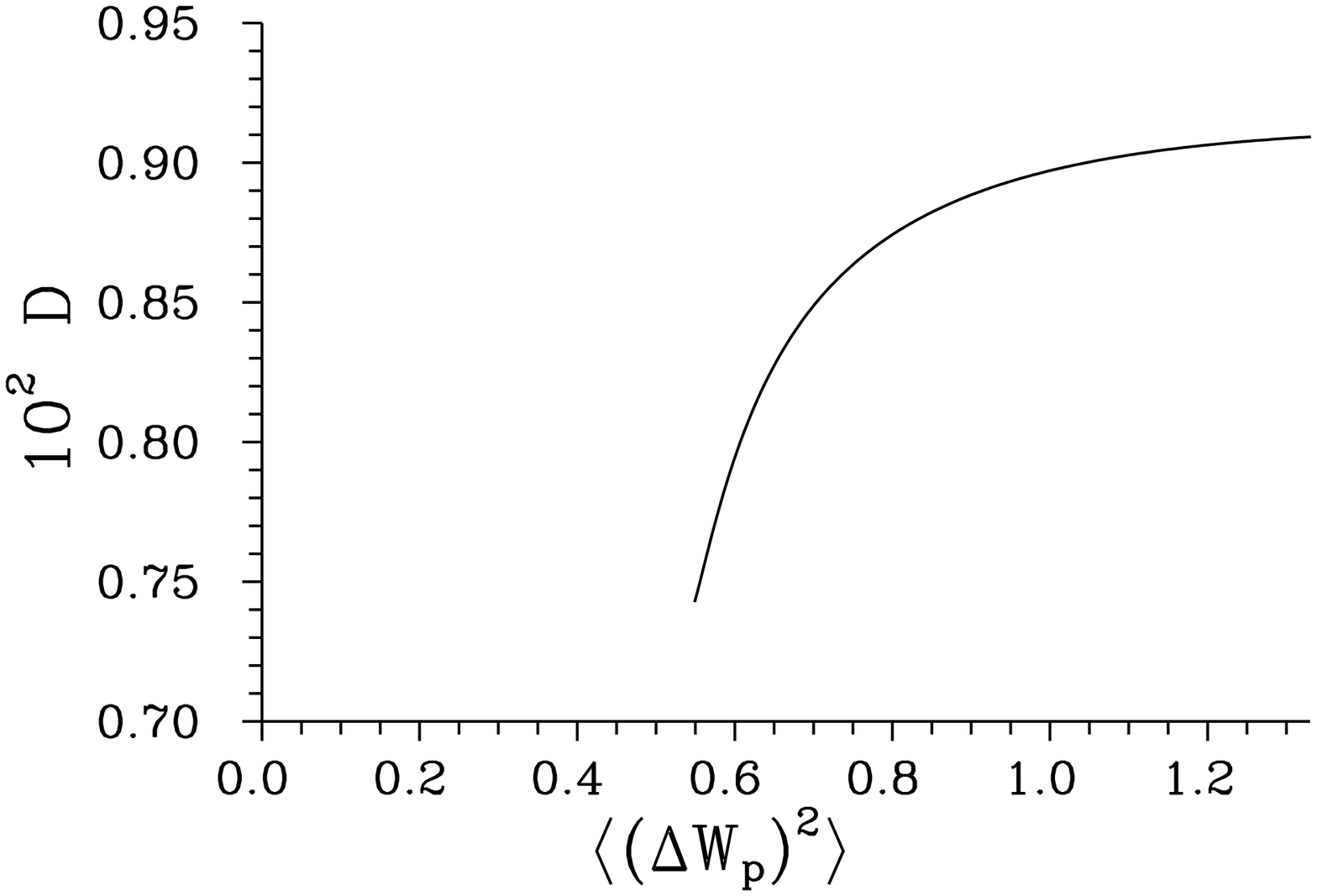}}

 \vspace{5mm}
 (b)
 \resizebox{0.8\hsize}{!}{\includegraphics[scale=0.26]{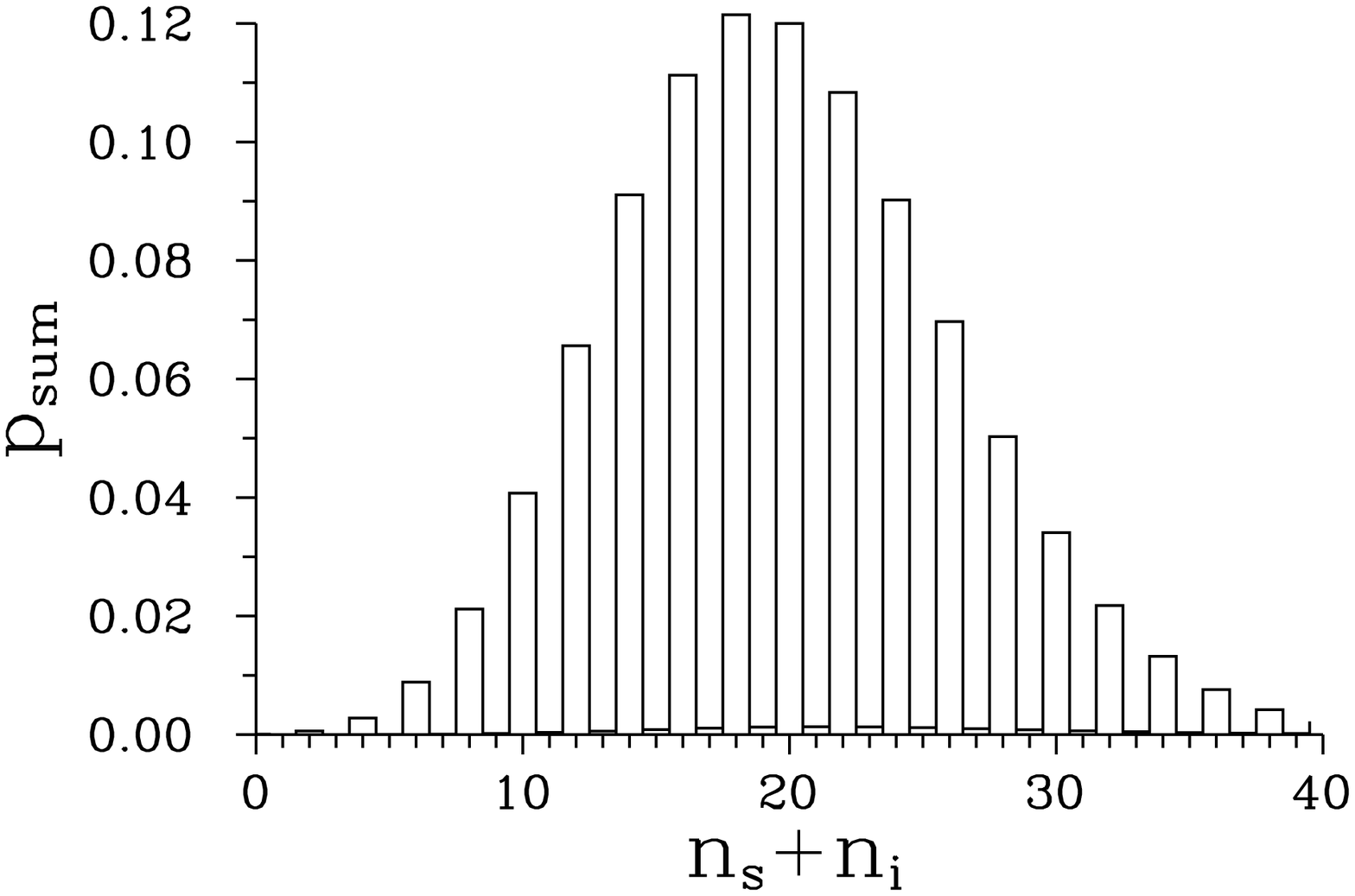}}

 \caption{(a) Function $ {\cal D} $ giving the declination of theoretical
 photocount distribution and experimental photocount histogram as a function of
 moment $ \langle (\Delta W_p)^2 \rangle $. Function $ {\cal D} $ is defined only for
 the values of $ \langle (\Delta W_p)^2 \rangle $ fulfilling Eq.~(\ref{4}). (b)
 Distribution $ p_{\rm sum} $ of the sum $ n_s + n_i $ of the signal and idler photon
 numbers for $ \langle (\Delta W_p)^2 \rangle = 0.549 $.}
\label{fig1}
\end{figure}

The dependence of function $ {\cal D} $ on the 'last free'
parameter $ \langle (\Delta W_p)^2 \rangle $ of the investigated
state attains a global minimum as shown in Fig.~\ref{fig1}(a).
This minimum is reached for $ \langle (\Delta W_p)^2 \rangle =
0.549 $ at the border of the allowed values. In this point, the
separation into paired and signal/idler single-photon noise fields
is such that the numbers of modes and their mean photon numbers
attain the values: $ M_p = 179 $, $ B_p = 0.055 $, $ M_s = 8
\times 10^{-6} $, $ B_s = 320 $, $ M_i = 8 \times 10^{-3} $ and $
B_i = 12 $. This means that the reconstructed field contains on
average 9.9 photon pairs and 0.003 (0.1) signal (idler) noise
photons. Thus, almost 99\% of the detected photoelectrons have
their origin in the detection of photon pairs.
\begin{figure}%[htb]    % fig. 2
 (a) \resizebox{0.8\hsize}{!}{\includegraphics{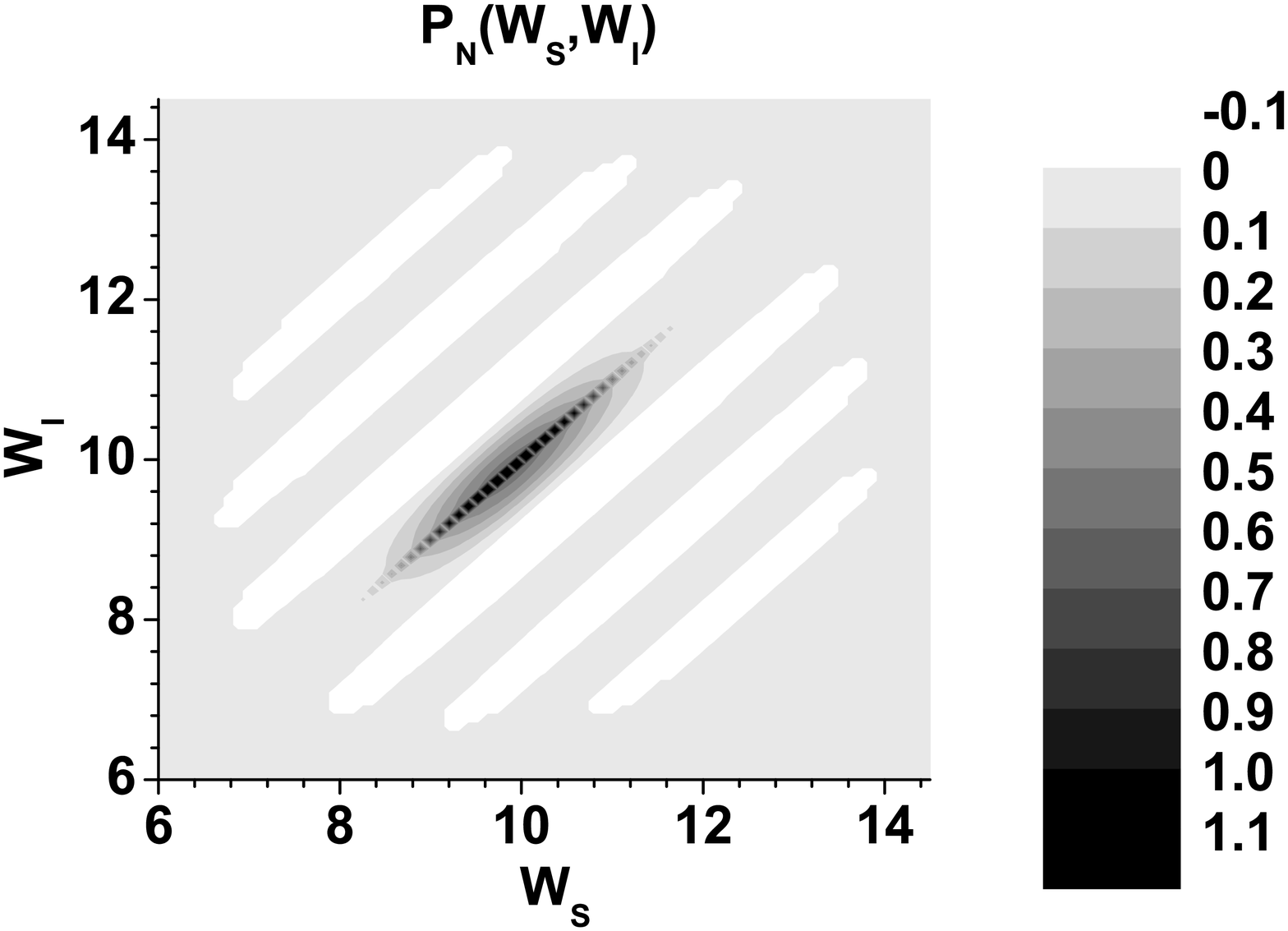}}

 \vspace{5mm}
 (b) \resizebox{0.8\hsize}{!}{\includegraphics{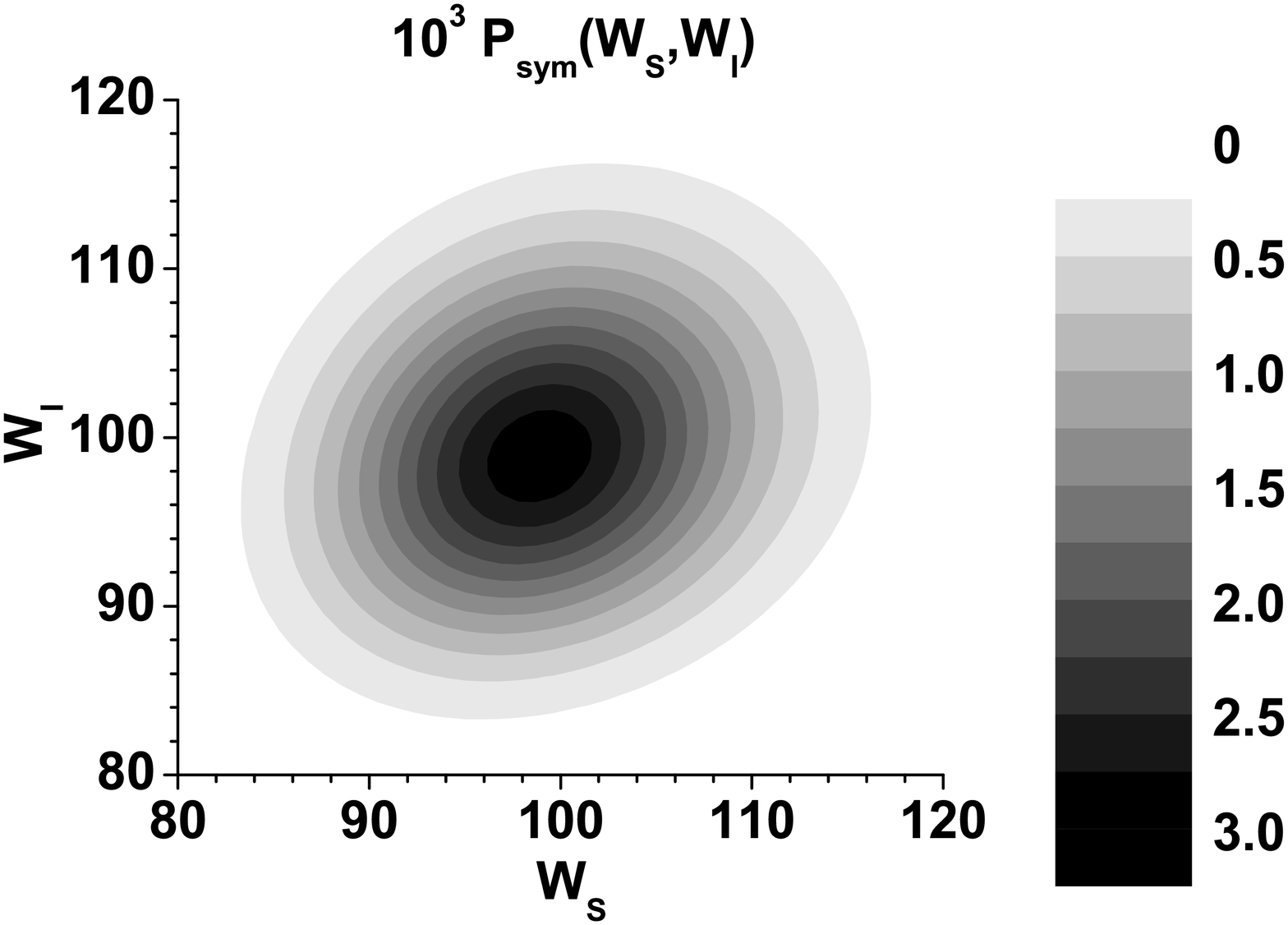}}
 \caption{Topo graphs of (a) normally-ordered ($ {\sf s}=1 $) and (b)
  symmetrically-ordered ($ {\sf s}=0 $) joint SI quasi-distributions
  $ P(W_s,W_i) $ of integrated intensities; $ {\sf s}_{\rm th} = 0.63 $.
  Integrated intensities $ W_s $ and $ W_i $ are in units of photon numbers.}
\label{fig2}
\end{figure}

\section{Quasi-distributions of integrated intensities}

According to Eq.~(1), QDII $ P(W_s,W_i) $ can be determined as the
inverse Fourier transform of the characteristic function $ C_W $
defined in Eq.~(\ref{2}) ($ {\sf s} $ means the ordering parameter
of field operators \cite{Perina1991}):
\begin{eqnarray}  % 12
 P_{\sf s}(W_s,W_i) \hspace{-2mm} &=& \hspace{-2mm} \int_{0}^{\infty}
  dW'_s \int_{0}^{\infty} dW'_i
  P_{p,{\sf s}}(W_s-W'_s,W_i-W'_i) \nonumber \\
 & & \mbox{} \times
  P_{s,{\sf s}}(W'_s) P_{i,{\sf s}}(W'_i).
\label{12}
\end{eqnarray}
In Eq.~(\ref{12}), QDII $ P_{p,{\sf s}} $ of the field of photon
pairs takes the form \cite{Perina2005}:
\begin{eqnarray}  % 13
 P_{p,{\sf s}}(W_s,W_i) &=& \frac{({W_s W_i})^{(M_p-1)/2}}{\pi\Gamma(M_p)
  B_{p,{\sf s}}^{M_p} } \exp\left( -\frac{W_s+W_i}{2B_{p,{\sf s}}} \right)
  \nonumber \\
 & & \hspace{-5mm} \mbox{} \times
  \sqrt{-K_{p,{\sf s}}} {\rm sinc} \left(
  \frac{W_s-W_i}{\sqrt{-K_{p,{\sf s}}}} \right) , \hspace{5mm}
  {\sf s} > {\sf s}_{\rm th} , \nonumber \\
 & & \hspace{-15mm} \mbox{} = \; \frac{({W_s W_i})^{(M_p-1)/2}}{\Gamma(M_p)
  K_{p,{\sf s}} D_p^{M_p-1}}
  \exp\left[ -\frac{B_{p,{\sf s}}(W_s+W_i)}{K_{p,{\sf s}}} \right]
  \nonumber \\
 & &  \hspace{-15mm} \mbox{} \times
  {\rm I}_{M_p-1} \left(2
  \frac{D_p\sqrt{W_sW_i}}{K_{p,{\sf s}}} \right) , \hspace{5mm}
  {\sf s} < {\sf s}_{\rm th} ;
\label{13}
\end{eqnarray}
$ B_{p,{\sf s}} = B_p + (1-{\sf s})/2 $, $ D_p = \sqrt{B_p(B_p+1)}
$, $ K_{p,{\sf s}} = -{\sf s} B_p + (1-{\sf s})^2/4 $ and  $ {\sf
s}_{\rm th} = 1+2(B_p-\sqrt{B_p(B_p+1)}) $. Symbol $ I_M $ denotes
the modified Bessel function and $ {\rm sinc}(x) = \sin(x)/x $. On
the other hand, QDIIs $ P_{a,{\sf s}} $ of the multi-thermal noise
fields occurring in the convolution in Eq.~(\ref{12}) and related
to $ {\sf s} $ ordering are derived as \cite{Perina1991}:
\begin{equation}  % 14
 P_{a,{\sf s}}(W_a) = \frac{W_a^{M_a-1}}{\Gamma(M_a)
  B_{a,{\sf s}}^{M_a} } \exp\left( -\frac{W_a}{B_{a,{\sf s}}} \right),
  \hspace{2mm} a=s,i;
\label{14}
\end{equation}
$ B_{a,{\sf s}} = B_a + (1-{\sf s})/2 $. Provided that the noise
multi-thermal fields are weak, the QDII $ P_{\sf s} $ in
Eq.~(\ref{12}) inherits the features of QDII $ P_{p,{\sf s}} $
characterizing the field of photon pairs. For the paired field and
according to Eq.~(\ref{13}), the QDII $ P_{p,{\sf s}} $ maintains
its quantum form provided that we are close to the normal-ordering
of field operators ($ {\sf s} > {\sf s}_{\rm th} $). It attains
negative values localized in strips parallel to the diagonal due
to the pairwise character of the SI field [see
Fig.~\ref{fig2}(a)]. Faithful description of these strips with
negative values requires the introduction of additional
independent parameters that complete the five parameters of a
general Gaussian form. From this point of view, the separation of
fully entangled quantum (paired) part from the rest of the field
is extraordinarily convenient as it introduces only one additional
independent parameter. For lower values of ordering parameter $
{\sf s} $ ($ {\sf s} < {\sf s}_{\rm th} $), the superimposed
detection noise described by the 'ordering of field operators'
conceals negative values and quantum features (entanglement
\cite{Perina2011}) of the SI field. This results in a non-negative
QDII $ P_{p,{\sf s}} $ with a smoothed shape [see
Fig.~\ref{fig2}(b)] that can be successfully approximated by a
general two-dimensional Gaussian distribution function. In this
case, the needed five independent parameters can naturally be
determined from the first and second experimental moments. This is
usual for any classical field. We note here that a QDII related to
any value of ordering operator $ {\sf s}_{\rm th} $ can be derived
from that related to normal ordering using convolution with an
appropriate Gaussian function \cite{Perina1991}. However, this
procedure cannot be inverted for principal reasons
\cite{Perina1991}. That is why, we need the QDIIs related to
normal ordering for full characterization of non-classical states
including twin beams.

Considering the general QDII $ P_{\sf s} $ in the form of
Eq.~(\ref{12}), the threshold value $ {\sf s}_{\rm th} $ can be
obtained from the analysis of the first and second moments of
integrated intensities related to an arbitrary $ {\sf s} $
ordering. The condition $ \langle [\Delta(W_s-W_i)]^2
\rangle_{{\sf s}_{\rm th}} =0 $ can be rearranged into the
formula:
\begin{eqnarray}  % 15
 {\sf s}_{\rm th} &=& 1+2\left(\beta-\sqrt{\gamma^2-\beta}
  \right) ,
\label{15} \\
 \beta &=& \frac{M_sB_s + M_iB_i + 2M_pB_p}{M_s+M_i+2M_p} ,
   \nonumber \\
 \gamma &=& \frac{M_sB_s^2 + M_iB_i^2 - 2M_pB_p}{M_s+M_i+2M_p}.
   \nonumber
\end{eqnarray}
Inspection of Eq.~({15}) shows that the threshold value $ {\sf
s}_{\rm th} $ is below 1 provided that $ M_sB_s^2 + M_iB_i^2 \le
2M_pB_p $. This condition for non-classicality of twin beams can
be rewritten as
\begin{equation}   % 16
 \langle (\Delta W_s)^2 \rangle + \langle (\Delta W_i)^2 \rangle
  < 2 \langle W_p \rangle.
\label{16}
\end{equation}
According to Eq.~(\ref{16}), single-photon noise in the signal and
idler fields has to be sufficiently compensated by the paired
field to keep non-classicality of twin beams.

Having a joint SI photon-number distribution $ p(n_s,n_i) $
written in Eq.~(\ref{7}), the distribution $ p_{\rm sum} $ of the
sum of signal and idler photon numbers with its characteristic
teeth-like structure (only even photon-numbers are present)
\cite{Waks2004} can easily be obtained [see Fig.~\ref{fig1}(b)
above]. It provides the simplest experimental evidence of the
presence of photon pairs in the analyzed field. The prevailing
paired structure of the SI field is also responsible for strong
sub-shot-noise correlations in the signal and idler photon-number
difference [$ \langle (n_s-n_i)^2 \rangle / (\langle n_s \rangle +
\langle n_i \rangle) = 1 + \langle (W_s-W_i)^2 \rangle / (\langle
W_s \rangle + \langle W_i \rangle) = 0.07 $]
\cite{Jedrkiewicz2004,Blanchet2008,Brida2009a}.

\section{Conclusions}

We have reconstructed a quasi-distribution of integrated
intensities of twin beams using the reliable first and second
experimental photocount moments and the method of least square
declinations. Whereas a general Gaussian form of the
quasi-distribution is suitable for a classical field, a more
general form is needed for quantum entangled states for which
negative values of the quasi-distribution are characteristic. The
consideration of quantum-classical transition has revealed a
momentum criterion of non-classicality of twin beams. We consider
the developed method robust and reliable and as such applicable
and prospective also in other areas of quantum physics.

\acknowledgments Support by projects P205/12/0382 of GA \v{C}R,
Operational Program Research and Development for Innovations -
European Regional Development Fund project CZ.1.05/2.1.00/03.0058
and Operational Program Education for Competitiveness - European
Social Fund project CZ.1.07/2.3.00/20.0058 of M\v{S}MT \v{C}R are
acknowledged. J.P.Jr. thanks J. Pe\v{r}ina for discussions.

\bibliography{perina}

\begin{thebibliography}{22}
\expandafter\ifx\csname natexlab\endcsname\relax\def\natexlab#1{#1}\fi
\expandafter\ifx\csname bibnamefont\endcsname\relax
  \def\bibnamefont#1{#1}\fi
\expandafter\ifx\csname bibfnamefont\endcsname\relax
  \def\bibfnamefont#1{#1}\fi
\expandafter\ifx\csname citenamefont\endcsname\relax
  \def\citenamefont#1{#1}\fi
\expandafter\ifx\csname url\endcsname\relax
  \def\url#1{\texttt{#1}}\fi
\expandafter\ifx\csname urlprefix\endcsname\relax\def\urlprefix{URL }\fi
\providecommand{\bibinfo}[2]{#2}
\providecommand{\eprint}[2][]{\url{#2}}

\bibitem[{\citenamefont{Dempster et~al.}(1977)\citenamefont{Dempster, Laird,
  and Rubin}}]{Dempster1977}
\bibinfo{author}{\bibfnamefont{A.~P.} \bibnamefont{Dempster}},
  \bibinfo{author}{\bibfnamefont{N.~M.} \bibnamefont{Laird}}, \bibnamefont{and}
  \bibinfo{author}{\bibfnamefont{D.~B.} \bibnamefont{Rubin}},
  \bibinfo{journal}{J. R. Statist. Soc. B} \textbf{\bibinfo{volume}{39}},
  \bibinfo{pages}{1} (\bibinfo{year}{1977}).

\bibitem[{\citenamefont{Vardi and Lee}(1993)}]{Vardi1993}
\bibinfo{author}{\bibfnamefont{Y.}~\bibnamefont{Vardi}} \bibnamefont{and}
  \bibinfo{author}{\bibfnamefont{D.}~\bibnamefont{Lee}}, \bibinfo{journal}{J.
  R. Statist. Soc. B} \textbf{\bibinfo{volume}{55}}, \bibinfo{pages}{569}
  (\bibinfo{year}{1993}).

\bibitem[{\citenamefont{Pe\v{r}ina}(1991)}]{Perina1991}
\bibinfo{author}{\bibfnamefont{J.}~\bibnamefont{Pe\v{r}ina}},
  \emph{\bibinfo{title}{Quantum Statistics of Linear and Nonlinear Optical
  Phenomena}} (\bibinfo{publisher}{Kluwer, Dordrecht}, \bibinfo{year}{1991}).

\bibitem[{Bou(2000)}]{Bouwmeester2000}
in \emph{\bibinfo{booktitle}{The Physics of Quantum Information}}, edited by
  \bibinfo{editor}{\bibfnamefont{D.}~\bibnamefont{Bouwmeester}},
  \bibinfo{editor}{\bibfnamefont{A.}~\bibnamefont{Ekert}}, \bibnamefont{and}
  \bibinfo{editor}{\bibfnamefont{A.}~\bibnamefont{Zeilinger}}
  (\bibinfo{publisher}{Springer, Berlin}, \bibinfo{year}{2000}).

\bibitem[{\citenamefont{Lvovsky and Raymer}(2009)}]{Lvovsky2009}
\bibinfo{author}{\bibfnamefont{A.~I.} \bibnamefont{Lvovsky}} \bibnamefont{and}
  \bibinfo{author}{\bibfnamefont{M.~G.} \bibnamefont{Raymer}},
  \bibinfo{journal}{Rev. Mod. Phys.} \textbf{\bibinfo{volume}{81}},
  \bibinfo{pages}{299–} (\bibinfo{year}{2009}).

\bibitem[{\citenamefont{Mandel and Wolf}(1995)}]{Mandel1995}
\bibinfo{author}{\bibfnamefont{L.}~\bibnamefont{Mandel}} \bibnamefont{and}
  \bibinfo{author}{\bibfnamefont{E.}~\bibnamefont{Wolf}},
  \emph{\bibinfo{title}{Optical Coherence and Quantum Optics}}
  (\bibinfo{publisher}{Cambridge Univ. Press, Cambridge},
  \bibinfo{year}{1995}).

\bibitem[{\citenamefont{Haderka et~al.}(2005)\citenamefont{Haderka,
  {Pe\v{r}ina~Jr.}, Hamar, and Pe\v{r}ina}}]{Haderka2005a}
\bibinfo{author}{\bibfnamefont{O.}~\bibnamefont{Haderka}},
  \bibinfo{author}{\bibfnamefont{J.}~\bibnamefont{{Pe\v{r}ina~Jr.}}},
  \bibinfo{author}{\bibfnamefont{M.}~\bibnamefont{Hamar}}, \bibnamefont{and}
  \bibinfo{author}{\bibfnamefont{J.}~\bibnamefont{Pe\v{r}ina}},
  \bibinfo{journal}{Phys. Rev. A} \textbf{\bibinfo{volume}{71}},
  \bibinfo{pages}{033815} (\bibinfo{year}{2005}).

\bibitem[{\citenamefont{{Pe\v{r}ina~Jr.}
  et~al.}(2012)\citenamefont{{Pe\v{r}ina~Jr.}, Hamar, Mich\'{a}lek, and
  Haderka}}]{PerinaJr2012}
\bibinfo{author}{\bibfnamefont{J.}~\bibnamefont{{Pe\v{r}ina~Jr.}}},
  \bibinfo{author}{\bibfnamefont{M.}~\bibnamefont{Hamar}},
  \bibinfo{author}{\bibfnamefont{V.}~\bibnamefont{Mich\'{a}lek}},
  \bibnamefont{and} \bibinfo{author}{\bibfnamefont{O.}~\bibnamefont{Haderka}},
  \bibinfo{journal}{Phys. Rev. A} \textbf{\bibinfo{volume}{85}},
  \bibinfo{pages}{023816} (\bibinfo{year}{2012}).

\bibitem[{\citenamefont{Haderka et~al.}(2004)\citenamefont{Haderka, Hamar, and
  {Pe\v{r}ina~Jr.}}}]{Haderka2004}
\bibinfo{author}{\bibfnamefont{O.}~\bibnamefont{Haderka}},
  \bibinfo{author}{\bibfnamefont{M.}~\bibnamefont{Hamar}}, \bibnamefont{and}
  \bibinfo{author}{\bibfnamefont{J.}~\bibnamefont{{Pe\v{r}ina~Jr.}}},
  \bibinfo{journal}{Eur. Phys. J. D} \textbf{\bibinfo{volume}{28}},
  \bibinfo{pages}{149} (\bibinfo{year}{2004}).

\bibitem[{\citenamefont{{\v{R}eh\'a\v{c}ek}
  et~al.}(2003)\citenamefont{{\v{R}eh\'a\v{c}ek}, Hradil, Haderka,
  {Pe\v{r}ina~Jr.}, and Hamar}}]{Rehacek2003}
\bibinfo{author}{\bibfnamefont{J.}~\bibnamefont{{\v{R}eh\'a\v{c}ek}}},
  \bibinfo{author}{\bibfnamefont{Z.}~\bibnamefont{Hradil}},
  \bibinfo{author}{\bibfnamefont{O.}~\bibnamefont{Haderka}},
  \bibinfo{author}{\bibfnamefont{J.}~\bibnamefont{{Pe\v{r}ina~Jr.}}},
  \bibnamefont{and} \bibinfo{author}{\bibfnamefont{M.}~\bibnamefont{Hamar}},
  \bibinfo{journal}{Phys. Rev. A} \textbf{\bibinfo{volume}{67}},
  \bibinfo{pages}{061801(R)} (\bibinfo{year}{2003}).

\bibitem[{\citenamefont{Ramilli et~al.}(2010)\citenamefont{Ramilli, Allevi,
  Chmill, Bondani, Caccia, and Andreoni}}]{Ramilli2010}
\bibinfo{author}{\bibfnamefont{M.}~\bibnamefont{Ramilli}},
  \bibinfo{author}{\bibfnamefont{A.}~\bibnamefont{Allevi}},
  \bibinfo{author}{\bibfnamefont{V.}~\bibnamefont{Chmill}},
  \bibinfo{author}{\bibfnamefont{M.}~\bibnamefont{Bondani}},
  \bibinfo{author}{\bibfnamefont{M.}~\bibnamefont{Caccia}}, \bibnamefont{and}
  \bibinfo{author}{\bibfnamefont{A.}~\bibnamefont{Andreoni}},
  \bibinfo{journal}{J. Opt. Soc. Am. B} \textbf{\bibinfo{volume}{27}},
  \bibinfo{pages}{852} (\bibinfo{year}{2010}).

\bibitem[{\citenamefont{Bondani et~al.}(2009)\citenamefont{Bondani, Allevi,
  Agliati, and Andreoni}}]{Bondani2009}
\bibinfo{author}{\bibfnamefont{M.}~\bibnamefont{Bondani}},
  \bibinfo{author}{\bibfnamefont{A.}~\bibnamefont{Allevi}},
  \bibinfo{author}{\bibfnamefont{A.}~\bibnamefont{Agliati}}, \bibnamefont{and}
  \bibinfo{author}{\bibfnamefont{A.}~\bibnamefont{Andreoni}},
  \bibinfo{journal}{J. Mod. Opt.} \textbf{\bibinfo{volume}{56}},
  \bibinfo{pages}{226} (\bibinfo{year}{2009}).

\bibitem[{\citenamefont{Allevi et~al.}(2010)\citenamefont{Allevi, Bondani, and
  Andreoni}}]{Allevi2010}
\bibinfo{author}{\bibfnamefont{A.}~\bibnamefont{Allevi}},
  \bibinfo{author}{\bibfnamefont{M.}~\bibnamefont{Bondani}}, \bibnamefont{and}
  \bibinfo{author}{\bibfnamefont{A.}~\bibnamefont{Andreoni}},
  \bibinfo{journal}{Opt. Lett.} \textbf{\bibinfo{volume}{35}},
  \bibinfo{pages}{1707} (\bibinfo{year}{2010}).

\bibitem[{\citenamefont{Miller et~al.}(2003)\citenamefont{Miller, Nam,
  Martinis, and Sergienko}}]{Miller2003}
\bibinfo{author}{\bibfnamefont{A.~J.} \bibnamefont{Miller}},
  \bibinfo{author}{\bibfnamefont{S.~W.} \bibnamefont{Nam}},
  \bibinfo{author}{\bibfnamefont{J.~M.} \bibnamefont{Martinis}},
  \bibnamefont{and} \bibinfo{author}{\bibfnamefont{A.~V.}
  \bibnamefont{Sergienko}}, \bibinfo{journal}{Appl. Phys. Lett.}
  \textbf{\bibinfo{volume}{83}}, \bibinfo{pages}{791} (\bibinfo{year}{2003}).

\bibitem[{\citenamefont{Pe\v{r}ina and K\v{r}epelka}(2011)}]{Perina2011}
\bibinfo{author}{\bibfnamefont{J.}~\bibnamefont{Pe\v{r}ina}} \bibnamefont{and}
  \bibinfo{author}{\bibfnamefont{J.}~\bibnamefont{K\v{r}epelka}},
  \bibinfo{journal}{Opt. Commun.} \textbf{\bibinfo{volume}{284}},
  \bibinfo{pages}{4941–} (\bibinfo{year}{2011}).

\bibitem[{\citenamefont{Leonhardt}(1997)}]{Leonhardt1997}
\bibinfo{author}{\bibfnamefont{U.}~\bibnamefont{Leonhardt}},
  \emph{\bibinfo{title}{Measuring the Quantum State of Light}}
  (\bibinfo{publisher}{Cambridge University Press, Cambridge},
  \bibinfo{year}{1997}).

\bibitem[{\citenamefont{Pe\v{r}ina and K\v{r}epelka}(2005)}]{Perina2005}
\bibinfo{author}{\bibfnamefont{J.}~\bibnamefont{Pe\v{r}ina}} \bibnamefont{and}
  \bibinfo{author}{\bibfnamefont{J.}~\bibnamefont{K\v{r}epelka}},
  \bibinfo{journal}{J. Opt. B: Quant. Semiclass. Opt.}
  \textbf{\bibinfo{volume}{7}}, \bibinfo{pages}{246} (\bibinfo{year}{2005}).

\bibitem[{\citenamefont{Pe\v{r}ina and K\v{r}epelka}(2006)}]{Perina2006}
\bibinfo{author}{\bibfnamefont{J.}~\bibnamefont{Pe\v{r}ina}} \bibnamefont{and}
  \bibinfo{author}{\bibfnamefont{J.}~\bibnamefont{K\v{r}epelka}},
  \bibinfo{journal}{Opt. Commun.} \textbf{\bibinfo{volume}{265}},
  \bibinfo{pages}{632} (\bibinfo{year}{2006}).

\bibitem[{\citenamefont{Waks et~al.}(2004)\citenamefont{Waks, Diamanti,
  Sanders, Bartlett, and Yamamoto}}]{Waks2004}
\bibinfo{author}{\bibfnamefont{E.}~\bibnamefont{Waks}},
  \bibinfo{author}{\bibfnamefont{E.}~\bibnamefont{Diamanti}},
  \bibinfo{author}{\bibfnamefont{B.~C.} \bibnamefont{Sanders}},
  \bibinfo{author}{\bibfnamefont{S.~D.} \bibnamefont{Bartlett}},
  \bibnamefont{and} \bibinfo{author}{\bibfnamefont{Y.}~\bibnamefont{Yamamoto}},
  \bibinfo{journal}{Phys. Rev. Lett.} \textbf{\bibinfo{volume}{92}},
  \bibinfo{pages}{113602} (\bibinfo{year}{2004}).

\bibitem[{\citenamefont{Jedrkiewicz et~al.}(2004)\citenamefont{Jedrkiewicz,
  Jiang, Brambilla, Gatti, Bache, Lugiato, and {Di~Trapani}}}]{Jedrkiewicz2004}
\bibinfo{author}{\bibfnamefont{O.}~\bibnamefont{Jedrkiewicz}},
  \bibinfo{author}{\bibfnamefont{Y.~K.} \bibnamefont{Jiang}},
  \bibinfo{author}{\bibfnamefont{E.}~\bibnamefont{Brambilla}},
  \bibinfo{author}{\bibfnamefont{A.}~\bibnamefont{Gatti}},
  \bibinfo{author}{\bibfnamefont{M.}~\bibnamefont{Bache}},
  \bibinfo{author}{\bibfnamefont{L.~A.} \bibnamefont{Lugiato}},
  \bibnamefont{and}
  \bibinfo{author}{\bibfnamefont{P.}~\bibnamefont{{Di~Trapani}}},
  \bibinfo{journal}{Phys. Rev. Lett.} \textbf{\bibinfo{volume}{93}},
  \bibinfo{pages}{243601} (\bibinfo{year}{2004}).

\bibitem[{\citenamefont{Blanchet et~al.}(2008)\citenamefont{Blanchet, Devaux,
  Furfaro, and Lantz}}]{Blanchet2008}
\bibinfo{author}{\bibfnamefont{J.-L.} \bibnamefont{Blanchet}},
  \bibinfo{author}{\bibfnamefont{F.}~\bibnamefont{Devaux}},
  \bibinfo{author}{\bibfnamefont{L.}~\bibnamefont{Furfaro}}, \bibnamefont{and}
  \bibinfo{author}{\bibfnamefont{E.}~\bibnamefont{Lantz}},
  \bibinfo{journal}{Phys. Rev. Lett.} \textbf{\bibinfo{volume}{101}},
  \bibinfo{pages}{233604} (\bibinfo{year}{2008}).

\bibitem[{\citenamefont{Brida et~al.}(2009)\citenamefont{Brida, Caspani, Gatti,
  Genovese, Meda, and Berchera}}]{Brida2009a}
\bibinfo{author}{\bibfnamefont{G.}~\bibnamefont{Brida}},
  \bibinfo{author}{\bibfnamefont{L.}~\bibnamefont{Caspani}},
  \bibinfo{author}{\bibfnamefont{A.}~\bibnamefont{Gatti}},
  \bibinfo{author}{\bibfnamefont{M.}~\bibnamefont{Genovese}},
  \bibinfo{author}{\bibfnamefont{A.}~\bibnamefont{Meda}}, \bibnamefont{and}
  \bibinfo{author}{\bibfnamefont{I.~R.} \bibnamefont{Berchera}},
  \bibinfo{journal}{Phys. Rev. Lett.} \textbf{\bibinfo{volume}{102}},
  \bibinfo{pages}{213602} (\bibinfo{year}{2009}).

\end{thebibliography}
\bibliographystyle{apsrev}

\end{document}